\documentclass[twocolumn,preprintnumbers,amsmath,amssymb,superscriptaddress]{revtex4}

\usepackage{graphicx}
\usepackage{epstopdf}
\usepackage{dcolumn}
\usepackage{bm}

\begin{document}

\title{Nuclear spin-lattice relaxation studies of Cu$_{2}$O$_{}$} 

\author{Yutaka Itoh\thanks{E-mail:yitoh@cc.kyoto-su.ac.jp}}
 \affiliation{Department of Physics, Graduate School of Science, Kyoto Sangyo University, Kamigamo-Motoyama, Kita-ku, Kyoto 603-8555, Japan}

\date{\today}

\begin{abstract}
We report the $^{63}$Cu and $^{65}$Cu nuclear spin-lattice relaxation rate measurements of cuprous oxide Cu$_2$O in a zero field Cu nuclear quadrupole resonance at $T$ = 77$-$325 K.
From the detailed isotopic measurements of the relaxation rates, we successfully estimated a finite magnetic relaxation rate $^{63}W_M$ and a predominant nuclear quadrupole relaxation rate $^{63}W_Q$.
$^{63}W_Q$ changed as $T^{2.1}$, while $^{63}W_M$ changed as $T^{1.6}$ or $T^{\beta}\mathrm{exp}(-\it\Delta/T)$ with $\beta$ = 0.6(3) and $\it\Delta$ = 190(62) K.
The nuclear spin scattering process due to a non-degenerate Fermi gas was discussed as a possible candidate of the magnetic relaxation. 
\end{abstract}



\maketitle

\section{Introduction}
Electron spin, charge, and lattice fluctuations play vital roles in solids,
because these fluctuations characterize the microscopic properties of the solids. 
Experimental efforts have been devoted to elucidate the individual fluctuations. 

Quadrupole nuclei can be powerful probes to detect magnetic and lattice fluctuations in solids. 
Naturally abundant quadrupole nuclei $^{63}$Cu and $^{65}$Cu have nuclear spin $I$ = 3/2 with different nuclear gyromagnetic ratios $^{63, 65}\gamma_n$ ($^{63}\gamma_n < {^{65}\gamma_n}$) and quadrupole moments $^{63, 65}Q$ ($^{63}Q > {^{65}Q}$)~\cite{NST}.
The nuclear spin-lattice relaxation can be due to magnetic relaxation via local magnetic field fluctuations and quadrupole relaxation via local electric field gradient fluctuations.
According to the general longitudinal nuclear spin relaxation theory, the magnetic nuclear spin-lattice relaxation rate $^{\eta}W_M$ is proportional to $({^{\eta}\gamma_n})^2$, while the quadrupole nuclear spin-lattice relaxation rate $^{\eta}W_Q$ is proportional to $({^{\eta}Q)^2}$ ($\eta$ = 63, 65 stands for $^{63}$Cu, $^{65}$Cu)~\cite{Abragam}.
The observed $^{\eta}$Cu nuclear spin-lattice relaxation rate in a solid of interest is denoted as $^{\eta}W_1$.
Both magnetic and quadrupole relaxations can contribute to $^{\eta}W_1$.   
It is an experimental criterion to identify the predominant relaxation process whether the isotopic ratio of the observed $^{\eta}$Cu nuclear spin-lattice relaxation rates is $^{63}W_1/^{65}W_1>$1 or $<$1. 
Furthermore, the accurate isotopic measurements of $^{63}W_1$ and $^{65}W_1$ may enable us to quantify how much role each relaxation plays in the nuclear spin-lattice relaxation.

Successful separation of mixed magnetic and quadrupole relaxation is found in isotopic nuclear quadrupole resonance (NQR) experiments of the chain-ladder composite compound Sr$_{14}$Cu$_{24}$O$_{41}$~\cite{Takigawa}, the high-$T_\mathrm{c}$ cuprate superconductor YBa$_2$Cu$_4$O$_8$~\cite{Mali}, 
the quasi-one dimensional conductor PrBa$_2$Cu$_4$O$_8$~\cite{Fujiyama}, 
and the kagome lattice Heisenberg antiferromagnet Zn-barlowite ZnCu$_3$(OD)$_6$FBr~\cite{Imai}.

Cu$_2$O has a cubic crystal structure with inversion symmetry.
The cuprous oxide is a direct band-gap semiconductor with monovalent Cu$^{+}$ and a large band gap of $\sim$2 eV~\cite{CO}. 
Excitons in the optical spectra have attracted much attention~\cite{Kittel}.
The exciton in Cu$_2$O is a bound state of a 4$s$-character electron and a 3$d$-character hole.  
The nature of Rydberg excitons and possible exciton Bose-Einstein condensation have been explored in Cu$_2$O~\cite{Rydberg,BEC}. 

Cu NQR experiments have been performed for Cu$_2$O.
The nuclear electric quadrupole spin relaxation is known to be predominate in the Cu nuclear spin-lattice relaxation of Cu$_2$O~\cite{JA}.
Two-phonon Raman scattering process accounts for the characteristic temperature dependence of the Cu spin-lattice relaxation rate 1/$T_1\propto T^2$~\cite{QR}.
However, the original data do not exclude the existence of a finite magnetic nuclear spin-lattice relaxation within experimental uncertainty~\cite{JA}.  
It remains to be understood how much a magnetic nuclear spin scattering process works in Cu$_2$O.


In this paper, we report the detailed $^{63,65}$Cu nuclear spin-lattice relaxation rates for a powder Cu$_2$O in a zero field Cu NQR at $T$ = 77$-$325 K. 
From the combination of the experimental isotopic relaxation rates, we found a finite magnetic relaxation rate $^{63}W_M$ separately from a predominant nuclear quadrupole relaxation rate $^{63}W_Q$. 
The temperature dependences of the separated relaxation rates were $^{63}W_Q\propto T^{2.1}$ and $^{63}W_M\propto{T^{1.6}}$ or $^{63}W_M\propto{T^{0.6}}\mathrm{exp}(-\it\Delta/T)$ with $\it\Delta$ = 190 K.  
 
\section{Experiments}
Zero-field $^{63, 65}$Cu NQR (nuclear spin $I$ = 3/2, magnetic quantum number $m$ = $\pm$3/2 $\leftrightarrow$ $\pm$1/2) experiments were carried out for the powder samples of commercially available Cu$_2$O (99.9 $\%$ purity from $Rare$ $metallic$ Co. Ltd.).
The powder sample was confirmed to be in single phase by powder X-ray diffraction patterns.   
A phase-coherent-type pulsed spectrometer was utilized to perform the Cu NQR experiments.  

The recovery curves of the $^{63, 65}$Cu nuclear magnetization were measured by recording the free-induction decay signal $F$($t$) following a sequence of $\pi$$-$$t$$-$$\pi/2$ pulses.
The experimental recovery curves $^{\eta}p(t)\equiv ^{\eta}F(\infty)$$-$$^{\eta}F(t)$ of $^{\eta}$Cu nuclear magnetization $^{\eta}F(t)$ ($\eta$ = 63, 65) (integrated free-induction decay) were analyzed by a single exponential function   
\begin{equation}
^{\eta}p(t) = {^{\eta}p(0)}{\rm exp}\left(- ^{\eta}W_1 t\right),
\label{eqFID}
\end{equation}  
where $^{\eta}p$(0) and a nuclear spin-lattice relaxation rate $^{\eta}W_1$ are the fitting parameters~\cite{Abragam}.  
 
\section{Experimental results} 
\subsection{Cu NQR spectrum and isotopic nuclear spin-lattice relaxation rates}
\begin{figure}[t]
 \begin{center} 
 \includegraphics[width=1.10\linewidth]{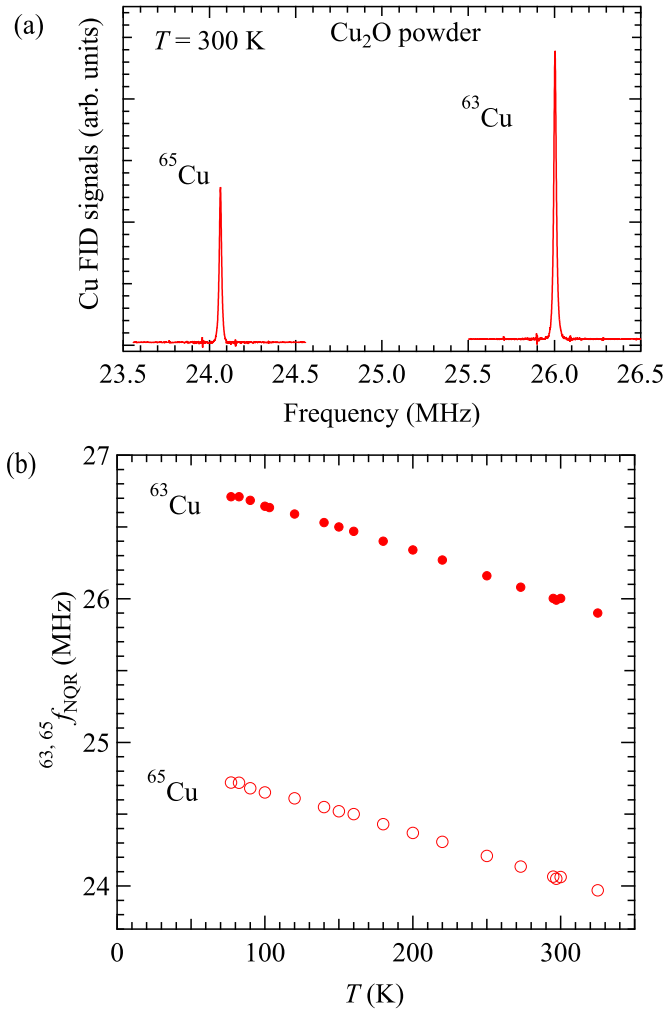}
 \end{center}
 \caption{\label{f1}(Color online)
(a) Fourier-transformed $^{63, 65}$Cu NQR frequency spectra of the free-induction decay signals ($m$ = $\pm$3/2 $\leftrightarrow$ $\pm$1/2) at 300 K. 
(b) $^{63, 65}$Cu NQR frequencies $^{63, 65}f_\mathrm{NQR}$ against temperature $T$. 
}
 \end{figure}
 
\begin{figure}[h]
 \begin{center} 
 \includegraphics[width=0.90\linewidth]{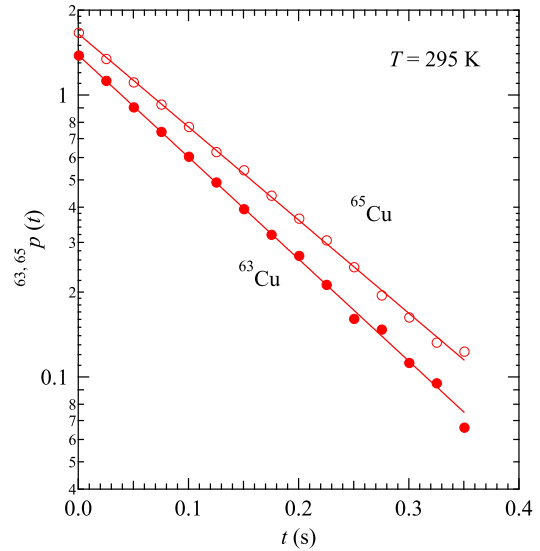}
 \end{center}
 \caption{\label{f2}(Color online)
Recovery curves $^{63, 65}p$($t$) of the $^{63, 65}$Cu nuclear free-induction decay signals at $T$ = 295 K in the Cu NQR. 
Solid curves are the least-squares fitting results using Eq.~(\ref{eqFID}).  
 }
 \end{figure}  
 
Figure~\ref{f1} (a) shows the Fourier-transformed $^{63, 65}$Cu NQR frequency spectra of the free-induction decay signals at 300 K.
The ratio of the NQR frequencies is that of the Cu nuclear quadrupole moments~\cite{CF}. 
We obtained
\begin{equation}
\frac{^{63}f_{\mathrm {NQR}}}{^{65}f_{\mathrm {NQR}}}=\frac{^{63}Q}{^{65}Q}=1.08061(3),
\label{Qr}
\end{equation}
which is consistent with the reported value~\cite{ZP}.

Figure~\ref{f1} (b) shows the $^{63, 65}$Cu NQR frequencies $^{63, 65}f_\mathrm{NQR}$ plotted against temperature $T$.
They are consistent with the reported $T$ dependences~\cite{nQT}. 

Figure~\ref{f2} shows the recovery curves $^{63, 65}p$($t$) of the $^{63,65}$Cu nuclear free-induction decay signals at $T$ = 295 K in the Cu NQR. 
The solid lines are the least-squares fitting results using Eq.~(\ref{eqFID}). 
 
\begin{figure}[t]
 \begin{center}
  \includegraphics[width=1.10\linewidth]{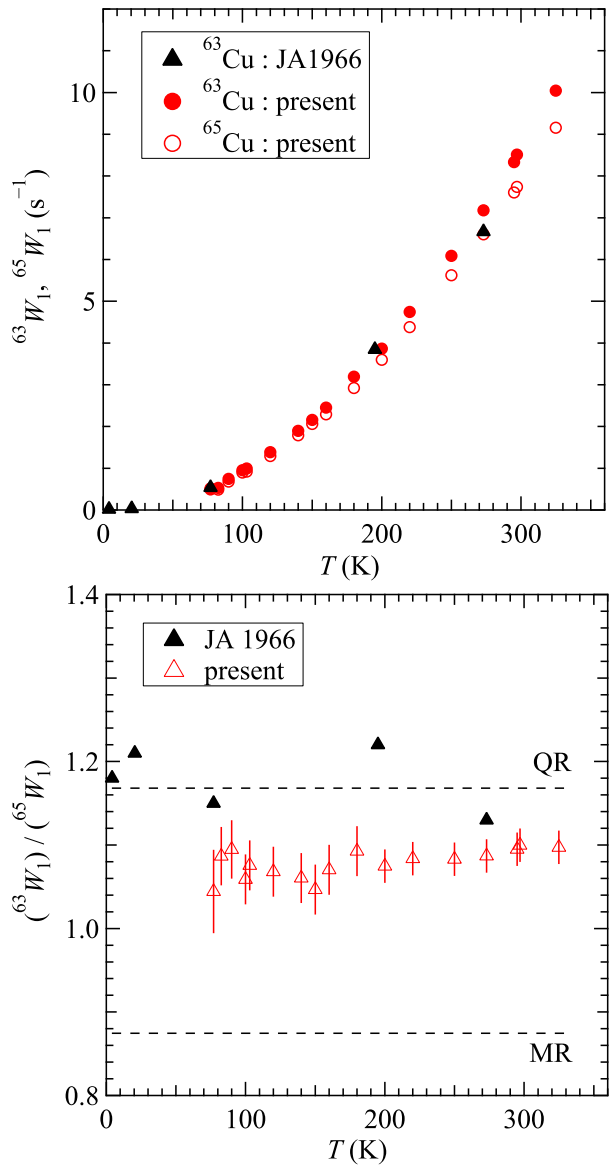}
 \end{center}
 \caption{\label{f3}(Color online)
Isotopic $^{63, 65}$Cu nuclear spin-lattice relaxation rates $^{63, 65}W_1$ against temperature $T$ (upper panel). The ratio $^{63}W_1/^{65}W_1$ against $T$ (lower panel). 
``JA1966" are the data (closed triangles) reproduced from Ref.~\cite{JA}. 
In the lower panel, QR stands for a pure quadrupole relaxation (the relaxation ratio of 1.168), and MR stands for a pure magnetic relaxation (the relaxation ratio of 0.8714).   
 }
  \end{figure}

The upper panel in Fig.~\ref{f3} shows the isotopic $^{63, 65}$Cu nuclear spin-lattice relaxation rates $^{63, 65}W_1$ against temperature $T$. 
The lower panel shows the ratios $^{63}W_1/^{65}W_1$ plotted against $T$. 
The closed triangles are the previous data (``JA1966") adopt from Ref.~\cite{JA}. 
In the lower panel, the present ratios of $^{63}W_1/^{65}W_1$ are close to but slightly smaller than $(^{63}Q/^{65}Q)^2$ = 1.168 for a pure quadrupole relaxation (QR) and larger than $(^{63}\gamma_n/^{65}\gamma_n)^2$ = 0.8714 for a pure magnetic relaxation (MR). 
We found a finite deviation from the pure quadrupole relaxation ratio by more accurate measurements than the JA1966 measurements.
 
\subsection{Quadrupole and magnetic relaxation rates}

\begin{figure}[t]
 \begin{center}
 \includegraphics[width=1.10\linewidth]{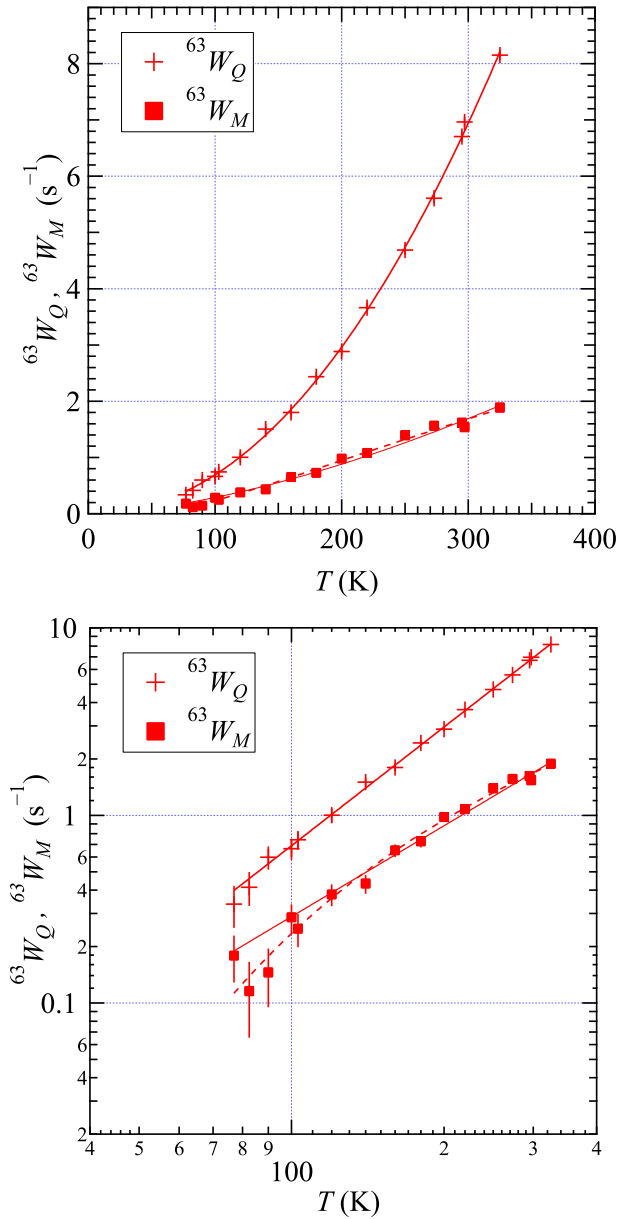}
 \end{center}
 \caption{\label{f4}(Color online)
Separated magnetic $^{63}$Cu nuclear spin-lattice relaxation rate $^{63}W_M$ and nuclear electric quadrupole spin-lattice relaxation rate $^{63}W_Q$ against temperature $T$ (upper panel) and versus $T$ in log-log plots (lower panel). 
Solid curves are the least-squares fitting results using a power law of Eq.~(\ref{eq7}) for $^{63}W_Q$ and $^{63}W_M$.  
Dashed curves are the least-squares fitting results using Eq.~(\ref{eq8}) for $^{63}W_M$.  
 }
 \end{figure}
  
In a zero field Cu NQR ($m$ = $\pm$3/2 $\leftrightarrow$ $\pm$1/2), 
the theoretical recovery curve with mixed magnetic and quadrupole relaxation is a single exponential function with a time constant $T_1$~\cite{Abragam}.
The magnetic dipole transition probability with $\Delta m$ = $\pm$1, the quadrupole transition probabilities with $\Delta m$ = $\pm$1 and $\pm$2 are additive in the single relaxation rate $W_1\equiv$1/$T_1$.
Thus, the experimental relaxation rates $^{\eta}W_1$ are expressed by the sum of the nuclear magnetic relaxation rate $^{\eta}W_M$ and the nuclear quadrupole relaxation rate $^{\eta}W_Q$ as 
\begin{eqnarray} 
^{63}W_1 &=& ^{63}W_M + ^{63}W_Q,\\  
^{65}W_1 &=& M_r ^{63}W_M + Q_r ^{63}W_Q, 
\label{W1}
\end{eqnarray}
where the coefficients $M_r$ = $(^{65}\gamma_n/^{63}\gamma_n)^2$ and $Q_r$ = $(^{65}Q/^{63}Q)^2$ act as conversion factors.

Using the above two equations, the quadrupole component $^{63}W_Q$ and the magnetic component $^{63}W_M$ in the $^{63}$Cu nuclear spin-lattice relaxation are expressed by the experimental $^{63, 65}W_1$ as 
\begin{eqnarray} 
^{63}W_Q &=& {1 \over {M_r - Q_r}}[M_r^{63}W_1 - ^{65}W_1],\\  
^{63}W_M &=& {1 \over {M_r - Q_r}}[^{65}W_1 - Q_r ^{63}W_1],
\label{WQM}
\end{eqnarray}
where $M_r$ = 1.1475 and $Q_r$ = 0.85639. 
Substitution of the experimental values for $^{63, 65}W_1$ in Eqs. (5) and (6) leads to $^{63}W_Q$ and $^{63}W_M$.

The upper panel in Fig.~\ref{f4} shows the magnetic $^{63}$Cu nuclear spin-lattice relaxation rate $^{63}W_M$ and the nuclear electric quadrupole spin-lattice relaxation rate $^{63}W_Q$ against temperature $T$. 
The lower panel shows the log-log plots of $^{63}W_M$ and $^{63}W_Q$ against $T$. 
The temperature dependence of $^{63}W_Q$ is of a power law type, while that of $^{63}W_M$ is somewhat different from the power law of $^{63}W_Q$. 
 
\section{Discussions}
The solid curves in Fig.~\ref{f4} are the least-squares fitting results for $^{63}W_Q$ and $^{63}W_M$ using a power law as
\begin{equation}
^{63}W_{Q,M} = aT^{b},
\label{eq7}
\end{equation}
where $a$ and $b$ are the fitting parameters. 

From the least-squares fitting result using Eq.~(\ref{eq7}), we obtained $b$ = 2.1(1) for $^{63}W_Q$.
This exponent of $^{63}W_Q$ is consistent with the original result and points to the two-phonon Raman scattering~\cite{JA,QR}.  
The nuclear spin scattering process due to lattice vibrations is the primary mechanism in the fluctuations of electric field gradients.  

From the least-squares fitting result using Eq.~(\ref{eq7}), we obtained $b$ = 1.6(1) for $^{63}W_M$. 
The exponent $b$ = 1.6(1) of $^{63}W_M$ is the same as that of the $^{29}$Si nuclear spin-lattice relaxation rate 1/$T_1\propto T^{1.6}$ in doped Si:P semiconductors, which is analyzed by a non-degenerate Fermi gas model~\cite{Warren}.
The mechanism in $^{63}W_M$ is due to dilute electron and hole scatterings in semiconductors~\cite{Abragam}. 
Theoretically, the degenerate Fermi gas at $T < T_\mathrm{F}$ (the Fermi temperature) yields the Korringa law, while the non-degenerate Fermi gas at $T > T_\mathrm{F}$ yields 1/$T_1\propto \sqrt{T}$ due to the electron density of states $\rho(E)\propto \sqrt{E}$~\cite{sqrtT,Abragam}. 
The Korringa law is insensitive to the shape of the density of states $\rho(E)$ but the power of $\sqrt{T}$ depends on the shape of $\rho(E)$ (1/$T_1\propto T^n$ for $\rho(E)\propto E^n$ with an arbitrary number $n$.)~\cite{Abragam}. 

A product function with fitting parameters ($\alpha$, $\beta$ and $\it\Delta$) of 
\begin{equation}
^{63}W_M = {\alpha}T^{\beta}e^{-{\it\Delta}/T}
\label{eq8}
\end{equation}
was also found to reproduce the experimental $^{63}W_M$ against $T$ as shown by the dashed curves in Fig.~\ref{f4}.
Equation~(\ref{eq8}) includes Eq.~(\ref{eq7}). 
From the least-squares fitting result using Eq.~(\ref{eq8}), we obtained the exponent $\beta$ = 0.6(3) and $\it\Delta$ = 190(62) K. 
The chi-square $\chi^2$ distribution using Eq.~(\ref{eq8}) was slightly smaller than that using the single power law of Eq.~(\ref{eq7}).  
The exponent $\beta$ = 0.6(3) is close to 1/$T_1\propto \sqrt{T}$ for the non-degenerate Fermi gas with $\rho(E)\propto \sqrt{E}$.
Then, the $\it\Delta$ is a crossover temperature,
at which the effect of a thermal activation process changes. 

Let us assume that $\it\Delta$ is an energy gap in the non-degenerate Fermi gas.  
Excitons may play a role at lower temperatures $T<\it\Delta$. 
Excitons are electrically neutral, so that they carry no charge. 
The spin states consist of the singlet state (paraexciton) and the triplet state (orthoexciton).
Since the ortho-to-paraexciton transition can flip a nuclear spin, 
we expect that the ortho-to-paraexciton transition can contribute to the magnetic nuclear spin-lattice relaxation rate.  
The exchange split energy between the orthoexciton state and the paraexciton state is 12 meV~\cite{Kuwa},
which is the same order of magnitude as $\it\Delta$ = 19 meV. 
The magnetic energy gap $\it\Delta$ may be associated with the low exchange energy of the ortho-to-paraexcitons. 
The other energy scales like an exciton binding energy and the band gaps are far larger than $\it\Delta$~\cite{Rydberg,BEC}. 
In principle, the orthoexcitons can couple with electric field gradients but the effect of the excitons on the quadrupole relaxation may be no match for that of the lattice vibrations.    

We believe that the difference between the present data and the JA1966 is primarily due to experimental accuracy. 
However, we should point out a possibility that the vacancy concentration in our sample might be different from that in the previous sample in Ref.~\cite{JA}. 
Heat treatments may introduce some deficiency and then carriers into Cu$_2$O~\cite{CO}. 
An enhancement in $^{63}$(1/$T_1$) is reported in a low-temperature prepared Cu$_2$O~\cite{LP}. 
The magnitude of the magnetic relaxation rate $^{63}W_M$ may be associated with the carrier concentration in Cu$_2$O.              

\section{Conclusions}
In conclusion, we found a finite magnetic Cu nuclear spin-lattice relaxation in Cu$_2$O. 
The Cu nuclear spin-lattice relaxation in Cu$_2$O results from a predominate nuclear quadrupole relaxation and a small magnetic relaxation.
The separated nuclear quadrupole relaxation rate $^{63}W_Q\propto T^{2.1}$ can be due to lattice vibrations. 
The magnetic nuclear spin-lattice relaxation rate $^{63}W_M$ as a function of temperature was $^{63}W_M\propto T^{1.6}$ or $^{63}W_M\propto T^{\beta}\mathrm{exp}(-\it\Delta/T)$ with $\beta$ = 0.6(3) and $\it\Delta$ = 190(62) K. 
We discussed the power law, $\beta$ and $\it\Delta$ using a non-degenerate Fermi gas model.

\section*{ACKNOWLEDGMENT}
We thank M. Isobe (Max-Planck-Institut f\"{u}r Festk\"{o}rperforschung) for X-ray diffraction measurements.

\end{document}